\documentclass[journal]{IEEEtran}
\usepackage[T1]{fontenc}
\usepackage[utf8]{inputenc}
\usepackage{cite}
\usepackage{amsmath,amssymb,amsfonts}
\usepackage{algorithmic}
\usepackage{graphicx}
\graphicspath{{figures/}}
\usepackage{textcomp}
\usepackage{xcolor}
\usepackage{siunitx}  
\usepackage{caption}
\usepackage{subcaption}
\usepackage{tabularx}
\usepackage{eurosym}
\usepackage{cleveref}

\usepackage[font=footnotesize,labelfont=bf]{caption}

\def\BibTeX{{\rm B\kern-.05em{\sc i\kern-.025em b}\kern-.08em
    T\kern-.1667em\lower.7ex\hbox{E}\kern-.125emX}}

\begin{document}

\title{Development of a Low-Cost, Autonomous Pulse Amplitude Modulated (PAM) 
Chlorophyll Fluorometer for In-Situ Monitoring of Photosystem II Efficiency}

\author{Samaneh~Baghbani,
        Uygar~Akkoc,
        Clara~Stock,
        Christiane~Werner,
        and~Stefan~J.~Rupitsch
\thanks{S.~Baghbani, U.~Akkoc and S.~J.~Rupitsch are with the Department of Microsystems
Engineering (IMTEK), University of Freiburg, Freiburg, Germany
(e-mail: samaneh.baghbani@imtek.uni-freiburg.de; uygar.akkoc@imtek.uni-freiburg.de;
stefan.rupitsch@imtek.uni-freiburg.de).}%
\thanks{C.~Stock and C.~Werner are with the Department of Ecosystem Physiology, Institute of
Earth and Environmental Sciences, University of Freiburg, Freiburg, Germany
(e-mail: clara.stock@cep.uni-freiburg.de; c.werner@cep.uni-freiburg.de).}%
\thanks{Corresponding author: Samaneh Baghbani (e-mail: samaneh.baghbani@imtek.uni-freiburg.de).}%
\thanks{This work was supported by the German Research Foundation (DFG) within the
Collaborative Research Centre (CRC) 1537 -- ECOSENSE.}%
}

\maketitle

\begin{abstract}
The quantum yield efficiency of photosystem II ($\Phi_\text{PSII}$) is a parameter to assess the photosynthetic performance and stress status of plants. Commercial PAM fluorometers can measure this parameter, but they are expensive, bulky, or lack an autonomous operation. This contribution presents the development of an autonomous PAM fluorometer that overcomes these limitations and is suitable for large-scale deployment. It enables high spatio-temporal monitoring of $\Phi_\text{PSII}$ in forest canopies across a wide range of ambient light conditions. 
The prototype has a cost of approximately €150, geometric dimensions of \SI{3}{\centi\meter}~$\times$~\SI{6}{\centi\meter}~$\times$~\SI{2}{\centi\meter} and a weight of  \SI{50}{\gram}. In side-by-side tests across three plant species, it achieved measurement accuracy comparable to state-of-the-art commercial sensors, with a correlation factor of $R^2 = 0.95$.
\end{abstract}


\maketitle

\section{Introduction}
Photosynthesis is a fundamental process in plants that converts solar energy into chemical energy and serves as the primary energy source for all ecosystems. The light-dependent part of photosynthesis is driven by an array of pigment-protein complexes that capture and funnel excitation energy, mainly Photosystem II (PSII) and Photosystem I (PSI). They work in sequence to power electron transport and generate the chemical energy required for carbon fixation. The process begins when chlorophyll molecules absorb light and initiate a series of light-dependent reactions. In PSII, the absorbed energy is used to excite a molecule that releases an electron. This electron is passed along an electron transport chain to PSI, where it is energized again by additional light absorption. This energy is used to produce ATP, the main energy carrier of the cell, and NADPH, a molecule that stores high-energy electrons. ATP provides energy, while NADPH donates the electrons required to convert carbon dioxide into carbohydrates during the Calvin cycle \cite{taiz2015}.

Throughout this sequence, a small portion of the absorbed light energy is re-emitted as chlorophyll fluorescence, primarily from PSII. Chlorophyll fluorescence competes with photochemistry and with thermal energy dissipation (heat). Moreover, it provides a sensitive indicator of the functional status of PSII \cite{maxwell2000,Krause1991}. When PSII reaction centers are open and capable of electron transfer, fluorescence will be minimal. When they are closed due to saturation or stress, fluorescence will increase. This inverse relationship between fluorescence and photochemical efficiency makes chlorophyll fluorescence a powerful, non-invasive tool for monitoring photosynthetic performance and plant stress responses \cite{baker2008, murchie2013}.

One of the most informative parameters derived from chlorophyll fluorescence is the quantum yield efficiency of PSII ($\Phi_\text{PSII}$), which quantifies the proportion of absorbed light used in photochemistry under light-adapted conditions using \cite{genty1989}:
\begin{equation}
\Phi_{PSII} = \frac{F_m' - F}{F_m'},
\label{eq:quantum_yield}
\end{equation}
where $F$ denotes the steady state fluorescence yield and 
 $F_m'$ is the maximum fluorescence yield under light-adapted conditions. This parameter can be exploited to calculate the electron transport rate (ETR) through PSII, and the photosynthetic efficiency \cite{maxwell2000,schreiber1994}

\begin{equation}
ETR = \Phi_{PSII} \times PAR \times \alpha \times 0.5
\label{eq:etr}
\end{equation}
where $PAR$ is the incident photosynthetically active radiation, $\alpha$ is leaf absorptance, and 0.5 indicates that two photons are required to excite the photosystems, one photon each.  While $\Phi_{PSII}$ indicates the efficiency of light use, ETR shows how much total electron flow is occurring. Both parameters serve as early indicators of plant stress, often before visible symptoms arise \cite{Kalaji2014,stirbet2018}.

To measure chlorophyll a fluorescence, three main techniques are commonly used: (i) Fast Repetition Rate (FRR), (ii) OJIP transient method, and (iii) Pulse Amplitude Modulation (PAM) fluorometry \cite{Kalaji2014}. FRR fluorometry was originally developed for studying marine phytoplankton, and its optical geometry makes it suitable for liquid samples \cite{Kolber1998, suggett2003}. The OJIP method analyzes the rapid chlorophyll fluorescence transient induced by a strong actinic light following dark adaptation. It offers detailed insight into the PSII photochemical dynamics. However, it requires a dark adaptation time, which makes it impractical for continuous or autonomous use in the field site\cite{strasser2004,bates2019}. In contrast, PAM fluorometry remains the gold standard for terrestrial plant studies. It uses modulated measuring light to detect fluorescence under ambient light, which makes it ideal for real time, in situ monitoring of photosynthetic efficiency \cite{fu2022,lysenko2022,schreiber1986,schreiber2004}.

Tracking plant physiological responses has become very important as ecosystems face climate change. This creates a demand for an autonomous sensor network that can capture fine-scale variations in photosynthetic performance and enable ecosystem assessment \cite{porcar2008,porcar2011,Zhang2025}.

State-of-the-art PAM fluorometers developed by WALZ offer excellent measurement accuracy and are widely used in research\cite{Oivukkamaki2024,Linn2021,Rascher2000}. The MICRO‑PAM is a miniaturized fluorometer from Walz, that enables long-term, autonomous monitoring of plant photosynthetic efficiency without damaging the leaf \cite{walz_microPam_manual}. However, its high cost remains a barrier to large-scale deployment. In addition, the spatial resolution of Micro-PAM measurements is limited, as individual sensors must still be connected to a central data processing unit. Other commercial devices, such as those from LI-COR \cite{licor_li600} and Hansatech \cite{handypea+_brochure}, tend to be handheld, bulky, expensive, or poorly suited for long-term deployment due to physical contact with leaves or shading effects.

In recent years, in addition to advances in commercial instruments, several academic studies have focused on developing custom built PAM fluorometers. For example, Reimer et al. \cite{Reimer} developed an autonomous, wireless PAM fluorometer. However, the sensor performance was limited to measurements under dark or very low light conditions. Similarly, Haidekker et al.~\cite{haidekker2022lowcost} developed a low-cost PAM fluorometer using laser diodes as the excitation light source. Although this approach effectively reduced component costs, the emitted light intensity was insufficient to generate proper saturation pulses. As a result, their sensor could only be validated under low ambient light conditions. While it is promising in terms of affordability, the device's performance under higher light intensities limited its application in field site environments. 

A review of both commercial and academic PAM fluorometers reveals that, despite  recent advances, critical challenges remain unresolved. Particularly, the need to combine low cost, high measurement accuracy under variable ambient light, full autonomy, and long-term durability for field deployment. These issues restrict the widespread use of PAM fluorometry for high-resolution, continuous monitoring of photosynthetic efficiency in complex environments such as multi-layered forest canopies.

In this contribution, we present the design and performance evaluation of
a PAM fluorometer that could overcome these limitations and be ready for ``deploy-and-forget'' applications in natural ecosystems fulfilling the goals of the ECOSENSE project \cite{werner2024ecosense}. The device maintains high measurement accuracy across a wide range of ambient light conditions and is designed for scalable field deployment, with a projected unit cost below 150~\euro{}. It autonomously measures and records key chlorophyll fluorescence parameters, including steady-state fluorescence ($F$), maximum fluorescence ($F_m'$), and the quantum yield efficiency of photosystem II ($\Phi_{PSII}$). The developed PAM fluorometer transmits data wirelessly via LoRa (Long Range Radio) to a remote server. This architecture enables real-time, spatially distributed monitoring of plant physiological responses under field conditions.

\section{Materials and Methods}
\subsection{Principle of PAM Fluorometry and Saturation Pulse}
A significant challenge in measuring chlorophyll fluorescence under field conditions is to separate the weak fluorescence signal from strong ambient light \cite{Sonneveld1980}. The PAM technique addresses this by using short, modulated flashes of light—known as the \textit{measuring light}. Typically, the pulse duration is a few microseconds and has low frequency. Since the duration and the frequency of the pulses are small, its intensity can be relatively high without altering the physiological state of the leaf. As a result, the total number of photons reaching the leaf per second remains very low and causes no significant change in the degree of PSII openness \cite{Porcarcastell2014,baker2008}.

One of the most important aspects of this technique is its ability to capture key information about chlorophyll fluorescence. The measuring light has a constant intensity during the measurement, and variations in the detected fluorescence signal (spikes) occur because of changes in photosynthetic activity within the leaf \cite{schreiber2004,Porcarcastell2014}.

\begin{figure}[htbp]
  \centering
  \includegraphics[
    width=\linewidth,
    trim=0mm 60mm 0mm 120mm, clip
  ]{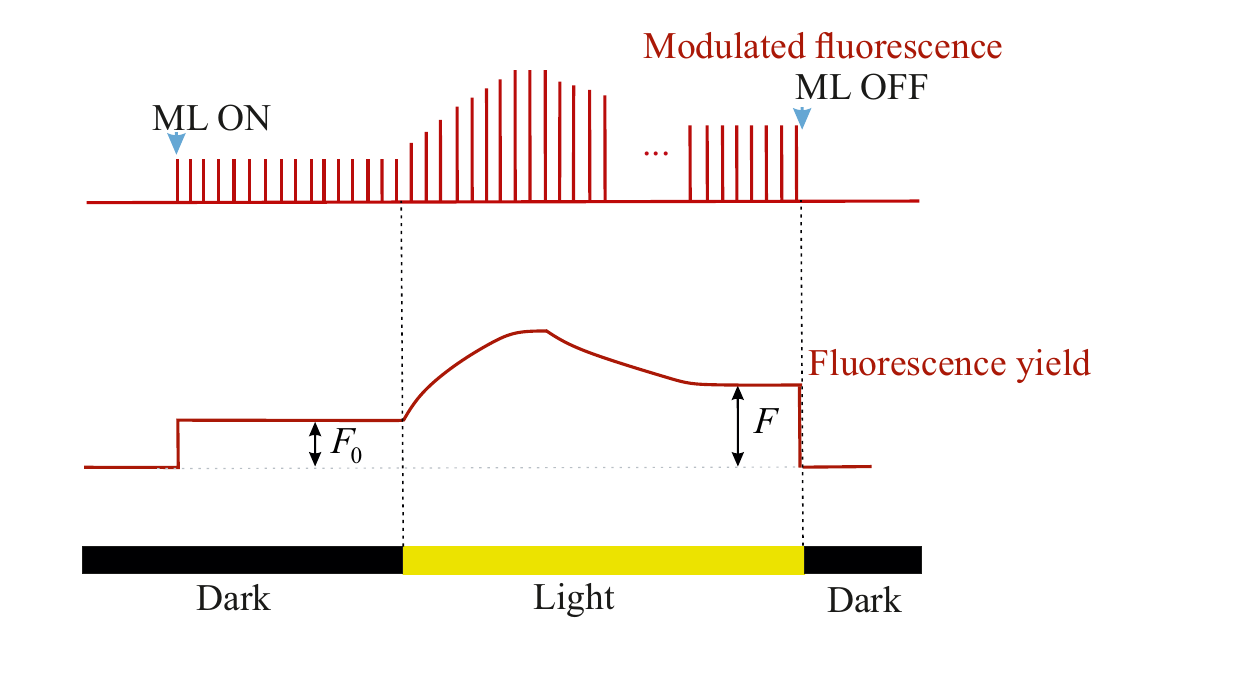} 
  \caption{Principle of PAM fluorometry within dark-light transition. The measuring light pulses are used to probe the photosynthetic activity of the leaf.}
  \label{fig:PAM_concept}
\end{figure}

Fig.~\ref{fig:PAM_concept} illustrates the principle of pulse-amplitude modulation fluorometry within a dark–light transition. When the measuring light is applied to a dark-adapted leaf, it produces the baseline fluorescence, denoted as $F_0$. In this state, the fluorescence yield is minimal because all PSII reaction centers are open, and most of the absorbed light energy is used for photochemistry rather than re-emitted as fluorescence.

When actinic light is turned on, the fluorescence detected by the measuring light pulses increases rapidly and reaches a transient maximum within a few hundred milliseconds because the absorbed light energy temporarily exceeds the capacity of the photosynthetic electron transport, and the excess energy is emitted as fluorescence. This peak typically lasts less than one second because the photosynthetic reactions and regulatory processes begin to activate and allow the plant to use the absorbed light for photochemistry. The fluorescence then decreases gradually as the plant’s photosynthetic processes become fully active and start to use more of the absorbed light energy. After several tens of seconds to a few minutes, light absorption and its utilization in photosynthesis become balanced and the fluorescence yield stabilizes at a steady-state level, $F$. This overall rise and fall in fluorescence signal upon exposure to continuous illumination is known as the Kautsky effect \cite{govindjee1995, bradbury1981, schreiber1986}.

Using the PAM technique with measuring light pulses, we can probe this dynamic even under strong illuminations. 

\begin{figure}[htbp]
  \centering
  \includegraphics[
    width=\linewidth,
    trim=10mm 100mm 0mm 50mm, clip
  ]{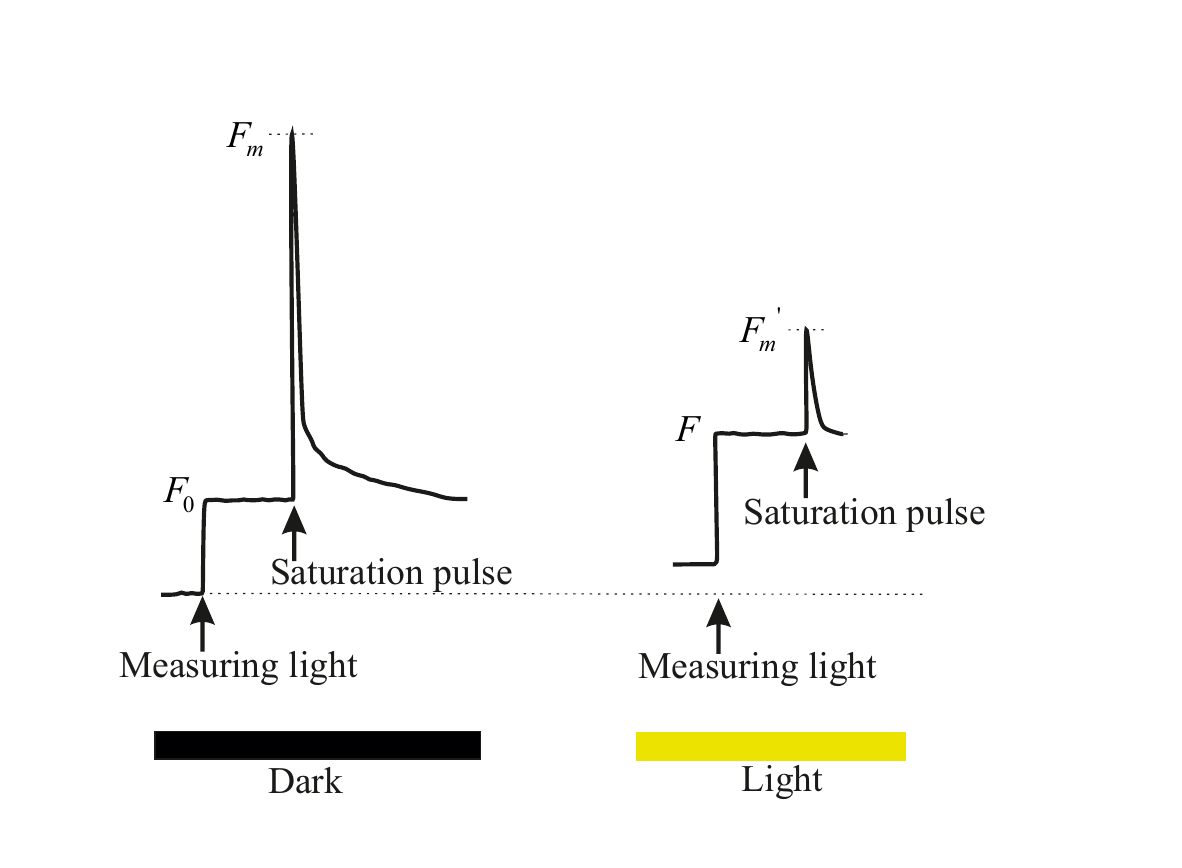} 
  \caption{Steady state fluorescence and maximum fluorescence in dark and light conditions.}
  \label{fig:F and Fm}
\end{figure}

In order to calculate $\Phi_\mathrm{PSII}$ under ambient light conditions, we need the steady-state fluorescence.
However, this value alone does not provide enough information to assess the plant’s stress level or photosynthetic efficiency. To have meaningful measurements, an additional light pulse called a \textit{saturation pulse} is applied. This is an intense flash of light—typically several thousand ~$\mu$mol m$^{-2}$ s$^{-1}$ for less than one second—that temporarily closes all PSII reaction centers, regardless of the ambient light level~\cite{schreiber2004, baker2008,Porcarcastell2014}. 
Fig.~\ref{fig:F and Fm} illustrates the steady state fluorescence yield and maximum fluorescence yield  in both dark and light-adapted leaves, respectively.

In a dark-adapted leaf, applying the saturation pulse produces the maximum fluorescence yield $F_m$, since all reaction centers are initially open and the pulse closes them completely. At this point, no absorbed energy can be used for photochemistry and is instead re-emitted as fluorescence \cite{Havaux1992}. For a non-stressed plant, the maximum quantum yield efficiency of PSII is usually observed under dark-adapted conditions and is approximately $\Phi_\mathrm{PSII,max} \approx 0.83$ \cite{Porcarcastell2014}.

When a saturation pulse is applied to a light-adapted leaf, the measured maximum fluorescence yield $F_\text{m}'$ will be usually lower than the $F_\text{m}$ value obtained from a dark adapted measurement. This is because, the plant is already using some of the absorbed light for photosynthesis, and mechanisms that dissipate excess energy as heat—known as non-photochemical quenching—are active. These processes reduce the amount of fluorescence emitted during the saturation pulse \cite{Karageorgou2007}. 

 Under a strong ambient light, many PSII centers are already partially closed, and fluorescence value $F$ is increased. Even when a high-intensity saturation pulse is applied, the additional increase in fluorescence value will be relatively small, and $F_\text{m}'$ is only slightly higher than $F$. As a result, the effective quantum yield of PSII, given in~\eqref{eq:quantum_yield}, decreases under high actinic light intensities.

In summary, a PAM fluorometer alternates between measuring lights and saturation pulses to calculate fluorescence parameters, enabling the calculation of quantum yield efficiency of PSII.

\subsection{Sensor Design and Architecture}

\begin{figure}[htbp]
  \centering
  \includegraphics[
    width=\linewidth,
    trim=10mm 130mm 0mm 40mm, clip
  ]{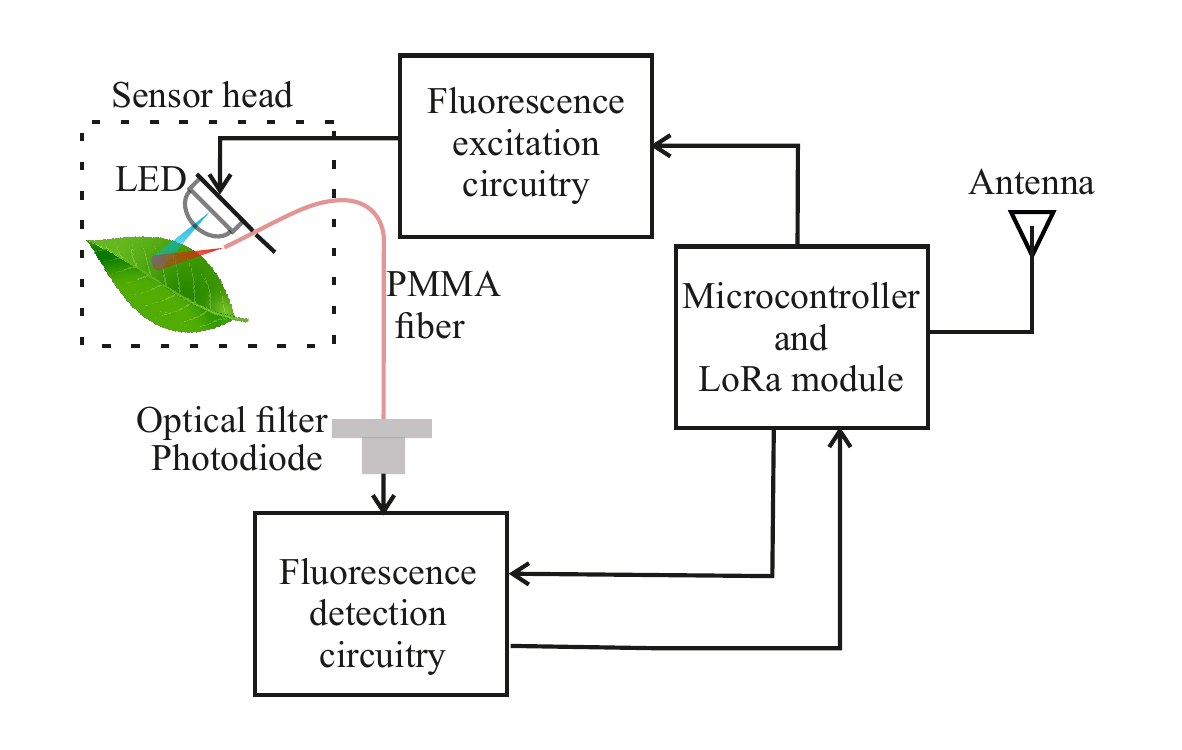} 
  \caption{Block diagram of the sensor system.}
  \label{fig:sensorblockdiagram}
\end{figure}

The developed sensor consists of four main parts: the sensor head, fluorescence excitation and detection circuitry, microcontroller, and LoRa communication module. As illustrated in Fig.~\ref{fig:sensorblockdiagram}, the sensor head attaches to the leaf and connects it to the sensor housing. The fluorescence excitation circuitry generates the excitation light and the fluorescence detection circuitry amplifies the resulting fluorescence signal. The microcontroller controls precisely all pulse timings during the measurement and processes the data to compute yield. Finally, the LoRa module transmits the measured parameters through an antenna to the gateway. Each part is described in detail below.

\subsubsection{Sensor Head} 

The sensor head is designed to hold the leaf in place and to facilitate both excitation and fluorescence signal collection. A blue LED with a peak emission wavelength of \SI{460}{\nano\meter} serves as the excitation light source and is mounted at a \SI{45}{\degree} angle to the leaf surface, at a distance of \SI{3}{\milli\meter}.

\begin{figure}[htbp]
  \centering
  \includegraphics[
    width=\linewidth,
    trim=10mm 90mm 0mm 100mm, clip
  ]{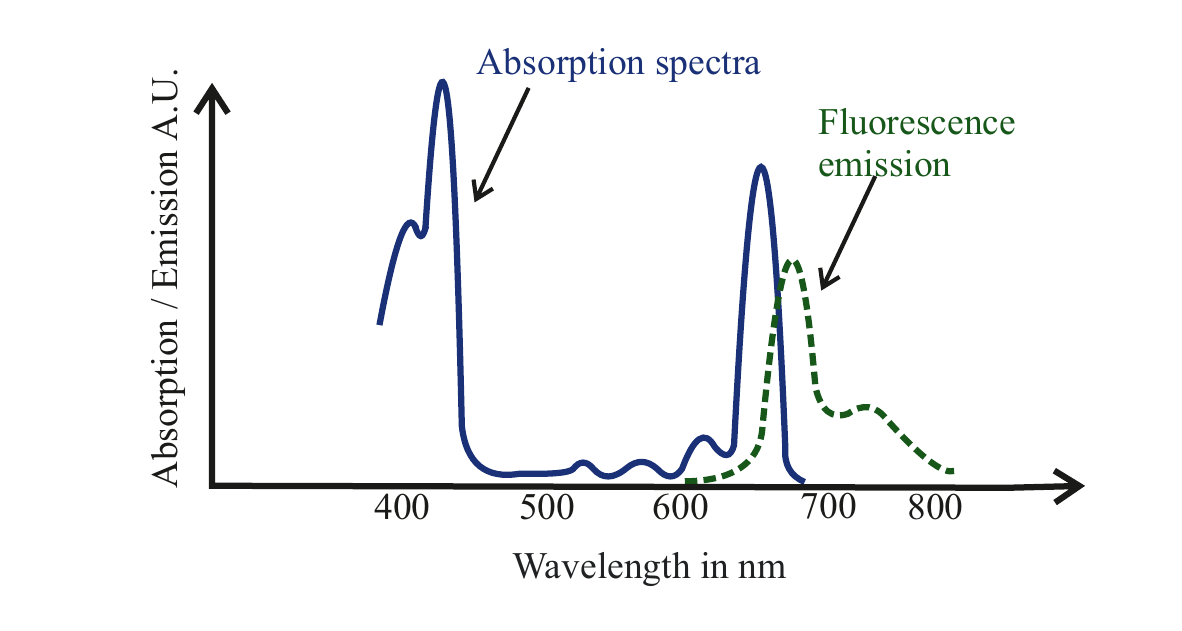} 
  \caption{Light absorption and emission by chlorophyll a molecules.}
  \label{fig:absorption_emission}
\end{figure}

To explain the choice of excitation source, it is important to recall that plants absorb light most efficiently in the blue and red regions of the spectrum, with absorption peaks around \SI{460}{\nano\meter} and \SI{640}{\nano\meter}, while fluorescence is emitted mainly in the red and far-red regions\cite{Porcarcastell2014}.
Fig.~\ref{fig:absorption_emission} illustrates the typical chlorophyll absorption and emission spectra.
Although both red and blue LEDs are suitable excitation sources, we selected blue light because this wider spectral gap between excitation and emission signals makes it easier to distinguish and isolate the fluorescence signal from the excitation background using one optical filter.

To collect the emitted chlorophyll fluorescence, a \SI{1.5}{\milli\meter} diameter PMMA plastic optical fiber is positioned adjacent to the LED, with a \SI{45}{\degree} angle to the leaf. This geometric arrangement ensures that the fluorescence is captured precisely from the area illuminated by the excitation light.
\subsubsection{Fluorescence Excitation}
As mentioned previously, excitation light in a PAM fluorometer consists of two sections: the measuring light and the saturation pulse. The measuring light is a series of modulated pulses used to probe chlorophyll fluorescence without significantly changing the photosynthetic state of the plant. In contrast, the saturation pulse is a brief, high-intensity light burst that momentarily closes all PSII reaction centers and enables the maximum fluorescence yield measurement.

In the sensor design, two different approaches can be used to generate the excitation light:

\textbf{Dual-LED Approach:} \\
One LED generates the measuring light and a separate LED delivers the saturation pulse, with each driven independently at different currents, pulse durations, and frequencies. 
When only the measuring light is applied, the variable fluorescence $F$ will be recorded. 
When the measuring light is applied simultaneously with the saturation pulse, the maximum fluorescence $F^{\prime}_{\mathrm{m}}$ will be obtained.

\textbf{Single-LED Approach:} \\
Alternatively, a single LED can be used to provide both the measuring light and the saturation pulse. 
In this approach, the LED operates initially with the parameters of the measuring light and then the saturation pulse is produced by driving the same LED at a very high duty cycle, creating nearly continuous high intensity light with only brief off times. 
These short off periods allow insertion of the measuring flashes between the high intensity bursts and enables $F^{\prime}_{\mathrm{m}}$ measurement without the need for a second LED.

\begin{figure}[htbp]
  \centering
  \includegraphics[
    width=\linewidth,
    trim=0mm 130mm 0mm 70mm, clip
  ]{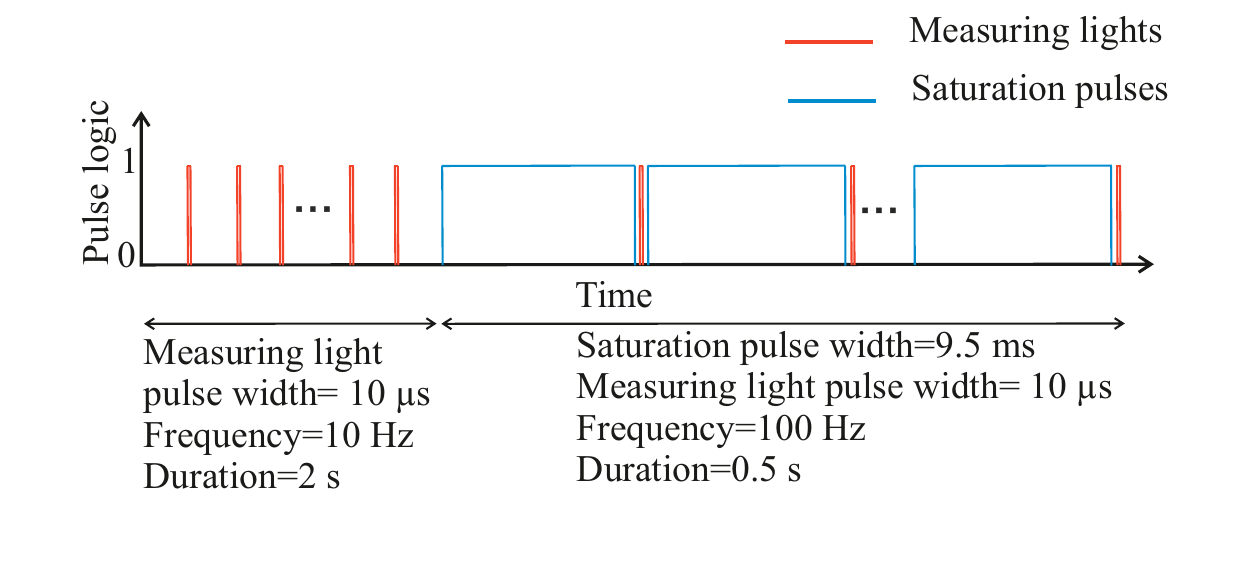} 
  \caption{Measurement protocol for the measuring light and saturation pulse.}
  \label{fig:ML and SP}
\end{figure}

We adopted the second approach because it makes the sensor more compact and simplifies the optical alignment. Fig.~\ref{fig:ML and SP} illustrates the timing sequence of the measuring light and saturation pulse used in this design. 
The measurement protocol begins with a \SI{2}{\second} application of measuring light at a frequency of \SI{10}{\hertz} and pulse duration of \SI{10}{\micro\second}. This measurement provides the $F$ value.
Then it is followed by \SI{0.5}{\second} of saturation pulses at \SI{100}{\hertz}, with an on-time of \SI{9.5}{\milli\second} for each pulse. The LED is turned off for \SI{500}{\micro\second} between saturation pulses. During this off-period, the measuring light is applied with the same intensity and pulse duration as before, starting \SI{200}{\micro\second} after the saturation pulse is turned off, also at \SI{100}{\hertz}. This arrangement ensures that the maximum fluorescence $F_m'$ is captured with high temporal precision. Since fluorescence begins to decline immediately once the saturation pulse terminates, 
a higher saturation pulse frequency helps maintain the leaf in a fully reduced state, and minimizes this decline and enables the observation of a clear saturation curve.

The LED is driven at two distinct current levels to generate the measuring light and the saturation pulse. Two LED drivers with enabling pins (BCR421U, Infineon Technologies), are used to supply the required currents for these two modes. The enable pins in LED drivers allow the microcontroller to precisely control the timing of both excitation signals. They deliver the required current with a rise time of less than \SI{500}{\nano\second}. Such a short rise time is essential to prevent the plant from adapting to the incident light and capturing fluorescence rise.

The continuous amplitude of the measuring light delivers a photosynthetic photon flux density (PPFD) of approximately \SI{1000}{\micro\mole\per\square\meter\per\second} at the leaf surface. 
Considering the total on time of the measuring pulses within one second, the LED is active for only about \SI{100}{\micro\second}, 
which corresponds to an effective PPFD of \SI{0.1}{\micro\mole\per\square\meter\per\second}. 
This very low average intensity is non actinic and does not alter the photosynthetic state of the leaf. Therefore, it 
ensures that the fluorescence signal reflects the true physiological condition rather than an artifact of the measuring light \cite{schreiber2004}. 

During the saturation pulse, the LED driver supplies a PPFD of approximately \SI{7000}{\micro\mole\per\square\meter\per\second}. Such a high PPFD is necessary to observe the saturation curve \cite{schreiber2004,maxwell2000}.

We applied measuring pulses \SI{200}{\micro\second} after the saturation pulse was turned off within the saturation pulse period. To determine this optimal time, we compared our measured yield values with those obtained using a commercial PAM fluorometer (Micro-PAM, Walz GmbH).
When measuring pulses were applied more than \SI{200}{\micro\second} after the saturation pulse ended, the calculated $\Phi_{PSII}$ especially under high ambient light was slightly lower than that obtained with the commercial device. 
Conversely, applying the measuring pulses earlier resulted in slightly higher yield values.

\subsubsection{Fluorescence  detection}

The fluorescence signal is collected by a PMMA optical fiber with a diameter of \SI{1.5}{\milli\meter} and guided to the sensor board, where it passes through a long-pass optical filter with a cutoff wavelength of \SI{650}{\nano\meter}. 
This filter blocks residual excitation light and transmits only the fluorescence emission to the detector.
A PIN photodiode is used as the detector due to its wide dynamic range and fast response time.

According to the PAM principle, the goal is to monitor changes in fluorescence upon applying the measuring light. 
While the measuring light remains constant, the fluorescence yield varies with actinic illumination and the physiological status of the leaf. 
By applying modulated measuring pulses, PAM enables tracking of fluorescence yield dynamics during dark-light transitions, under varying actinic lights, and during saturation pulses.

\begin{figure}[htbp]
  \centering
  \includegraphics[
    width=\linewidth,
    trim=0mm 180mm 0mm 30mm, clip
  ]{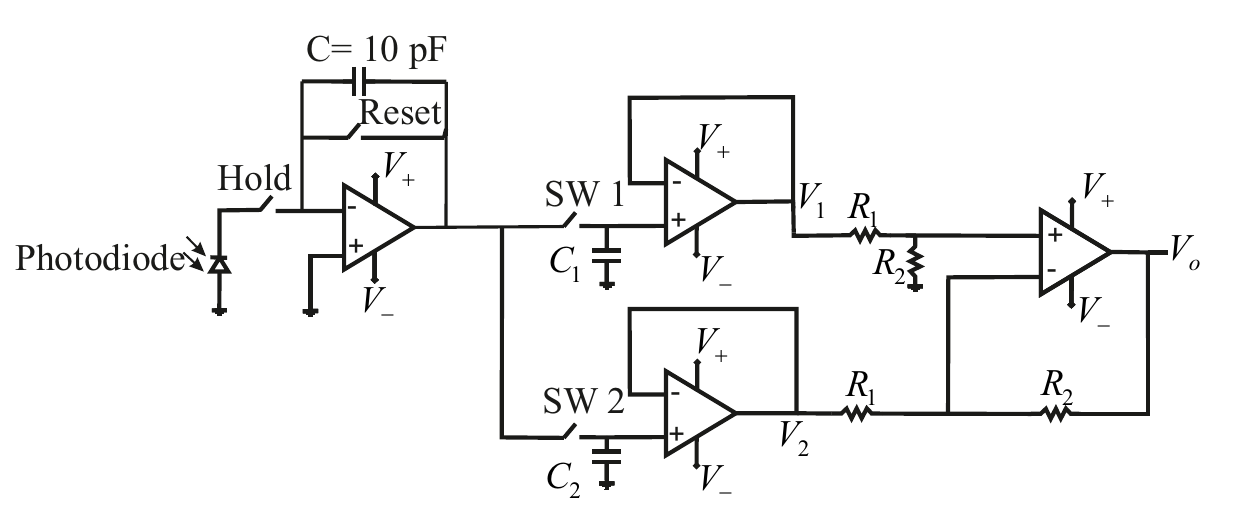}
  \caption{Detector circuitry for processing fluorescence yield signals.}
  \label{fig:detector circuitry}
\end{figure}

To implement this principle in hardware, we developed the detector circuitry shown in Fig.~\ref{fig:detector circuitry}. 
Its first stage is a switching integrator (IVC102, Texas Instruments) with a built-in \SI{10}{\pico\farad} capacitor, providing a gain of approximately $10^6$ during a \SI{10}{\micro\second} integration window. 
The IC includes two internal switches—reset and hold —that define the integration cycle: the reset discharges the capacitor to start a new integration window, while the hold switch allows current from the photodiode to flow into the integrator.

Each measurement cycle consists of two integration windows. 
The first window, taken immediately before the measuring pulse, captures fluorescence induced by ambient light. 
The second window which coincides with the measuring pulse, records the combined signal from ambient illumination and excitation. 
Switches SW1 and SW2 transfer these integrated signals to capacitors $C_1$ and $C_2$, which serve as sample and hold elements. 
The buffered signals are then passed to a differential stage, where subtraction isolates the fluorescence yield induced by the measuring light.

\begin{figure}[htbp]
  \centering
  \includegraphics[
    width=\linewidth,
    trim=0mm 60mm 0mm 70mm, clip
  ]{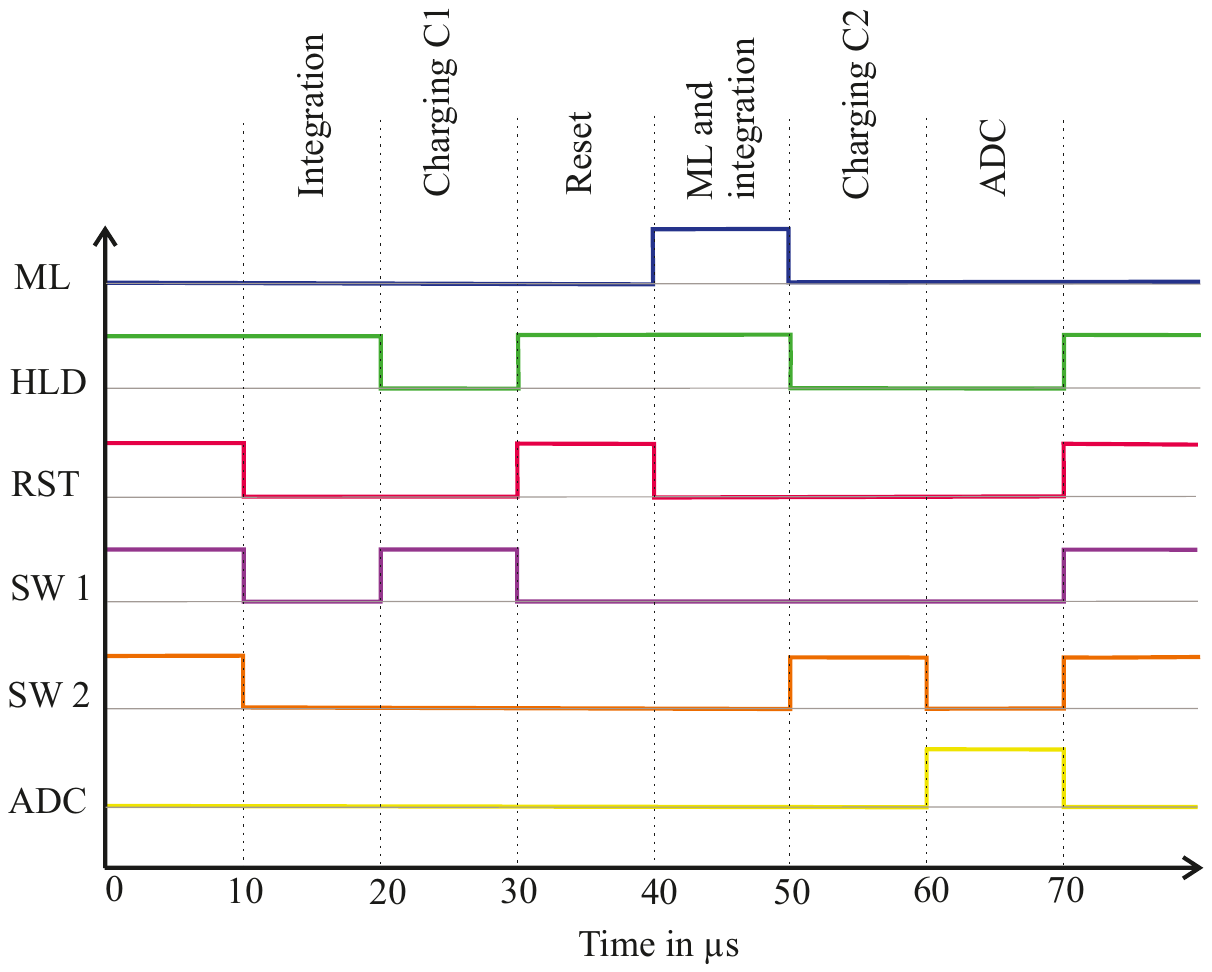} 
  \caption{Timing sequence of measuring light and switches in detector circuitry}
  \label{fig:timing sequence}
\end{figure}

Precise switch timing of the detector circuitry is illustrated in Fig.~\ref{fig:timing sequence}.
At the start of each cycle, all switches are closed to fully discharge the system and ground the photodiode. 
The first integration begins when SW1 and SW2 are opened to isolate $C_1$ and $C_2$, and the Reset switch is released while Hold remains closed, allowing the photodiode current to charge the integrator capacitor. 
At the end of this \SI{10}{\micro\second} window, the charge is transferred to $C_1$ and represents the fluorescence caused by ambient light. 
The process is then repeated in synchrony with the measuring light, and the second integrated value is stored on $C_2$. 
When both signals are stored, the differential stage will subtract and amplify the difference signal with an extra gain of 10. This signal then passes through a low-pass filter with a cut-off frequency of \SI{230}{\kilo\hertz} to suppress high frequency noise and prevent aliasing.

\subsection{Data Acquisition and Transmission}
The Apollo3 Blue microcontroller coordinates the entire measurement sequence, including the precise timing of the measuring light and saturation pulses, the control of the detector circuitry switches, analog-to-digital conversion (ADC), digital signal processing, and wireless data transmission via the LoRaWAN technology to a remote server.

To generate the measuring light pulses, the 48 MHz clock frequency of the microcontroller was used, while a 32 kHz external oscillator controlled the intervals between the measuring and saturation pulses. 

The conditioned fluorescence yield signal is digitized by the 12 bit ADC of the microcontroller. 
The ADC operates at a sampling frequency of \SI{1.2}{\mega\hertz} and is active only during the fluorescence yield acquisition windows. In order to minimize the power consumption, the ADC will remain inactive when no useful signal is present.
Within each \SI{10}{\micro\second} window, the differential signal is sampled twelve times, and the values are averaged to improve signal-to-noise ratio and suppress short-term fluctuations. 
This produces a single fluorescence yield value for each measuring pulse. 

To calculate $F$, the mean of 20 yield measurements is computed (corresponding to 2 seconds at \SI{10}{\hertz}). 
During the saturation phase, 50 yield  measurements are acquired (corresponding to 0.5 seconds at \SI{100}{\hertz}). The first 20 values are discarded to allow the leaf to reach full saturation, and the final 30 values are averaged to determine the $F_m'$ value. 
The quantum yield efficiency of Photosystem~II is then calculated according to~\ref{eq:quantum_yield}. 

Finally, the processed values $F$, $F_m'$, and $\Phi_\text{PSII}$ are transmitted via the integrated LoRa module to a gateway and forwarded to the server for storage and further analysis. The reason for choosing LoRa to transfer the data was its low cost, low power consumption, and wide coverage in forest environments \cite{mahjoub2024}.

\subsection{Sensor Cost Analysis}

Table~\ref{tab:components} lists the key components used to develop the sensor along with their sources and approximate prices. 
Additional parts such as resistors, capacitors, connectors and cables  are not listed individually due to their low cost. 
We further considered a total cost of 35~{\euro} for fabrication of the printed circuit board, 3D printing of the sensor head, housing and filter holder and PMMA fiber.
Including these items, the total cost of one complete sensor unit remains below {150~\euro}.

\begin{table}[!t]
\caption{List of components used in sensor development.}
\centering
\begin{tabularx}{\columnwidth}{|p{2cm}|X|p{2cm}|c|}
\hline
\textbf{Item} & \textbf{Description} & \textbf{Manufacturer} & \textbf{Cost (\euro)} \\
\hline
150141BS63130 & Blue LED 465 nm & Würth & 0.73 \\
\hline
15-203 & 650 nm Longpass optical filter & Edmund optics & 45.50 \\
\hline
BPW24R  & Photodiode & Vishay & 4.00 \\
\hline
BCR421U & LED driver & Infineon  & 0.77 \\
\hline
IVC102 & Switching Integrator & Texas Instruments & 9.50 \\
\hline
OPA2991 & Dual amplifier & Texas Instruments & 1.69 \\
\hline
TL071 & Op-amp Used as a difference amplifier. & Texas Instruments & 0.34 \\
\hline
LTC1563-2 & Low-pass filter & Analog Devices & 4.46 \\
\hline
TMUX1101 & CMOS Switch & Texas Instruments & 1.60 \\
\hline
AP2112K & LDO regulator 3.3V & Diodes Incorporated & 0.49 \\
\hline
LM340 & LDO regulator 5V & Texas Instruments & 1.31 \\
\hline
TDN-1 1222WI & DC/DC converter ±12V & TRACO Power & 18.69 \\
\hline
WRL-15484 & Ambiq Apollo 3 module & Ambiq & 8.80 \\
\hline
LAMBDA62-8S & LoRa module & RF Solutions & 13.48 \\
\hline
MB85RC64TAPNF -G-BDERE1 & 64kbit FeRAM & Ramxeed & 3.04 \\
\hline
AM1805AQ & Real time clock & Ambiq & 1.01 \\
\hline
SIP32431DR3-T1GE3 & Power switch ICs & Vishay & 0.76 \\
\hline
ABS06  & Crystal 32.768 kHz & ABRACON
 & 0.65 \\
\hline
\multicolumn{3}{|r|}{\textbf{Total}} & 114.91
 \\
\hline
\end{tabularx}
\label{tab:components}
\end{table}

\subsection{Sensor Housing}
\begin{figure}[htbp]
  \centering
  \includegraphics[
    width=\linewidth,
    trim=0mm 140mm 0mm 50mm, clip
  ]{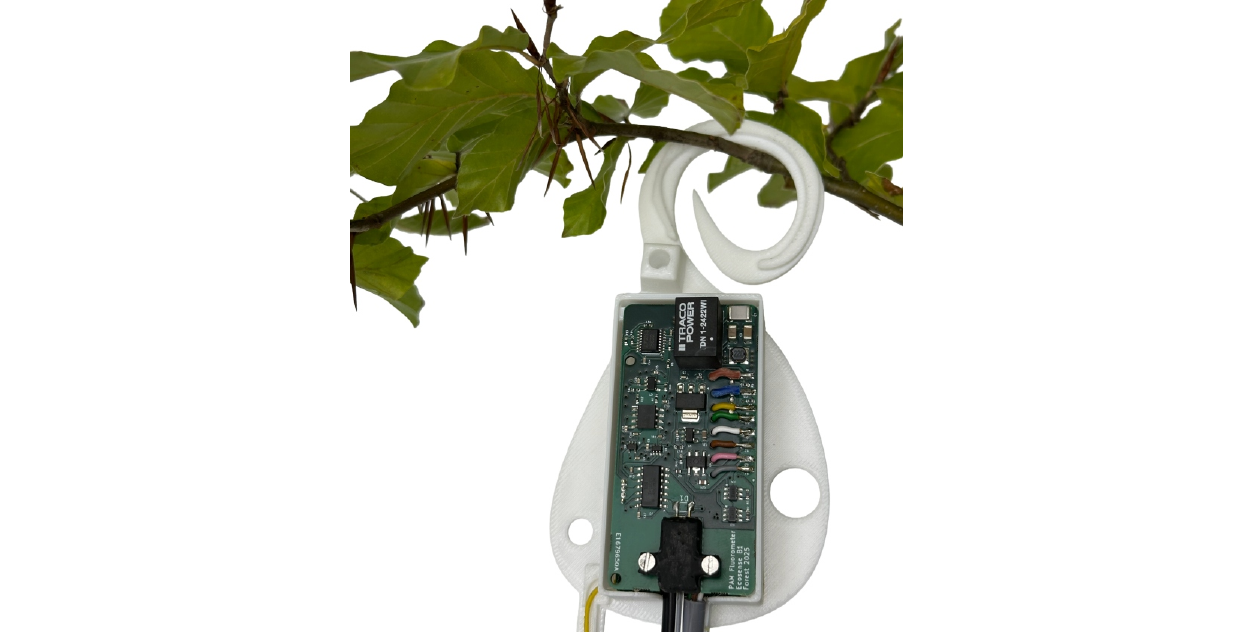} 
  \caption{Sensor housing designed specifically for hanging sensor in forest canopies.}
  \label{fig:Housing}
\end{figure}

The sensor housing was specifically designed to allow the sensor box to be hung from branches at various heights within forest canopies. Its hook enables easy attachment to branches, while the surrounding structure minimizes potential damage to nearby leaves and twigs. In addition, the input and output cables enter the housing from below to prevent water or rain from entering the enclosure.
Fig.~\ref{fig:Housing} illustrates how the sensor board is positioned inside the housing and attached to a branch. A protective lid closes the housing, and silicone glue is used to seal it for stable operation under forest conditions.

\section{Sensor Characterization and Validation}

To assess the accuracy and reliability of the developed sensor, we performed a two-stage validation. 
First, side-by-side measurements with a commercial PAM fluorometer were carried out to compare the performance of our device. 
Second, calibration and reproducibility tests were conducted to evaluate the stability and consistency of the measurements across replicates.

Experiments were carried out in a walk-in climate chamber (Thermo Tec) with regulated humidity, temperature, and lighting. 
The chamber provided light intensities with a sunlight like spectrum up to \SI{1400}{\micro\mole\per\square\meter\per\second}, depending on the distance of the objects from the ceiling light source. We performed tests across dark conditions and a range of light levels. 

For dark adapted conditions, measurements were conducted every hour, whereas for light adapted conditions, measurements were taken every 20 minutes. The timing between two measurements was chosen to allow the plant to recover from the previous measurement pulses, specially the saturation pulse, and to ensure that successive measurements did not interfere with each other.
The ambient light intensity was gradually increased stepwise up to the chamber’s maximum capability.

\subsection{Side-by-Side Measurements with a Commercial PAM Fluorometer}

To validate the performance of the developed sensor, side-by-side measurements were performed between our sensor and a commercial device (Micro-PAM, Walz GmbH) using leaves from three plant species: cherry laurel
(\textit{Prunus laurocerasus}), European beech
(\textit{Fagus sylvatica}), and sycamore maple
(\textit{Acer pseudoplatanus}). 
The $\Phi_\text{PSII}$ is highly sensitive to ambient light intensity, and even small variations in illumination can lead to significant differences in the measured values. To ensure that measurement points received identical illumination, we positioned both sensors on the same leaf, at the same height and angle, inside the chamber.
Since the developed sensor was not equipped with its own PAR sensor, the PAR values were obtained with the integrated PAR sensor of Micro-PAM. 

All light curve analyses presented in the results section are therefore based on the PAR data recorded by the Micro-PAM. At each light intensity, measurements were repeated at least 3 times to evaluate consistency.

\subsection{Reproducibility and Calibration}

To evaluate measurement reproducibility and further characterize the developed PAM fluorometers, a batch of 20 sensors was tested inside the climate chamber. 
The light intensity was gradually increased stepwise from darkness to \SI{1000}{\micro\mole\per\square\meter\per\second}, with each intensity level measured at least five times per sensor. 
All tests were performed using beech leaves as the sample plant. 
Each sensor was equipped with an LS-C mini quantum sensor (Walz GmbH) to record the corresponding PAR values.  

We quantified the reproducibility for each sensor by calculating the mean and standard deviation of the recorded $\Phi_{\text{PSII}}$ values at each light intensity. 
This experiment was designed to determine whether all sensors produced consistent and repeatable data across the tested light range and assess variations in measurement accuracy within the sensor batch, and finally to establish individual calibration factors where necessary.

\section{Results and Discussion}

\subsection{Results of Side-by-Side Measurements with the Commercial Device}

\begin{figure}[!t]
  \centering

  \begin{subfigure}[b]{0.9\linewidth}
    \centering
    \includegraphics[width=\linewidth, trim=0mm 70mm 20mm 80mm, clip]{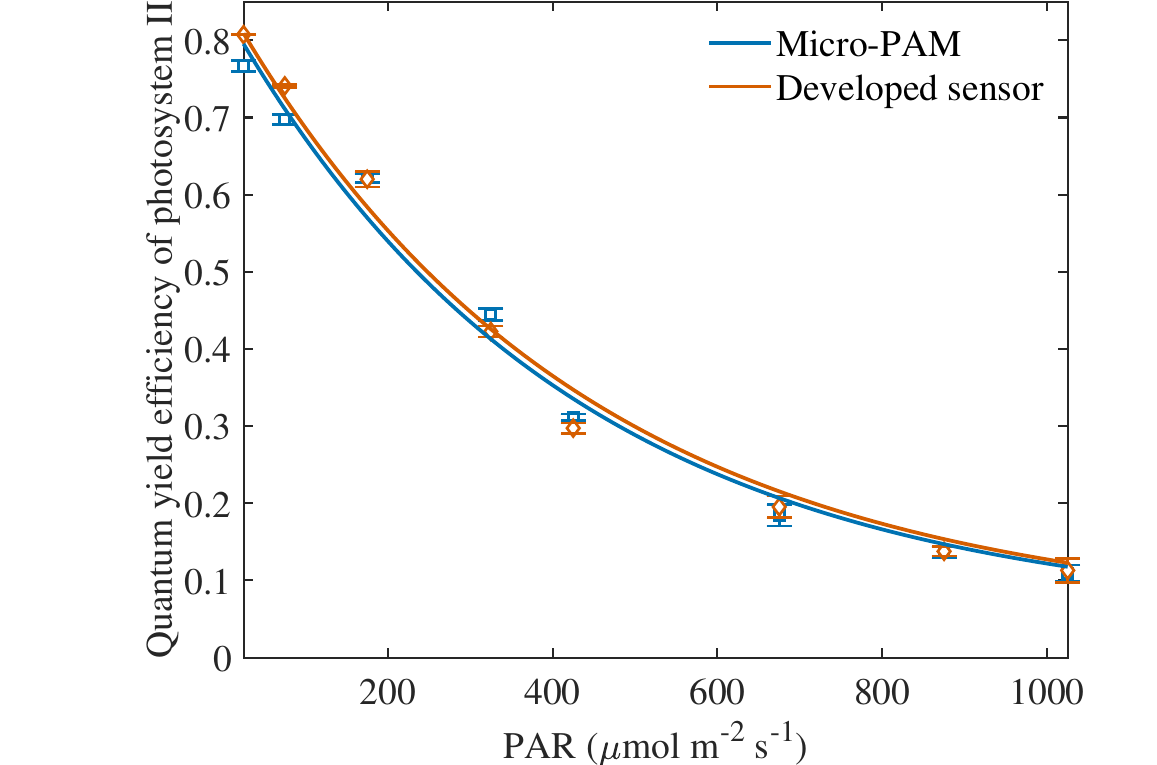}
    \caption{}
    \label{fig:Maple_Light_Curve2}
  \end{subfigure}

  \vspace{3mm} 

  \begin{subfigure}[b]{0.9\linewidth}
    \centering
    \includegraphics[width=\linewidth, trim=0mm 70mm 20mm 80mm, clip]{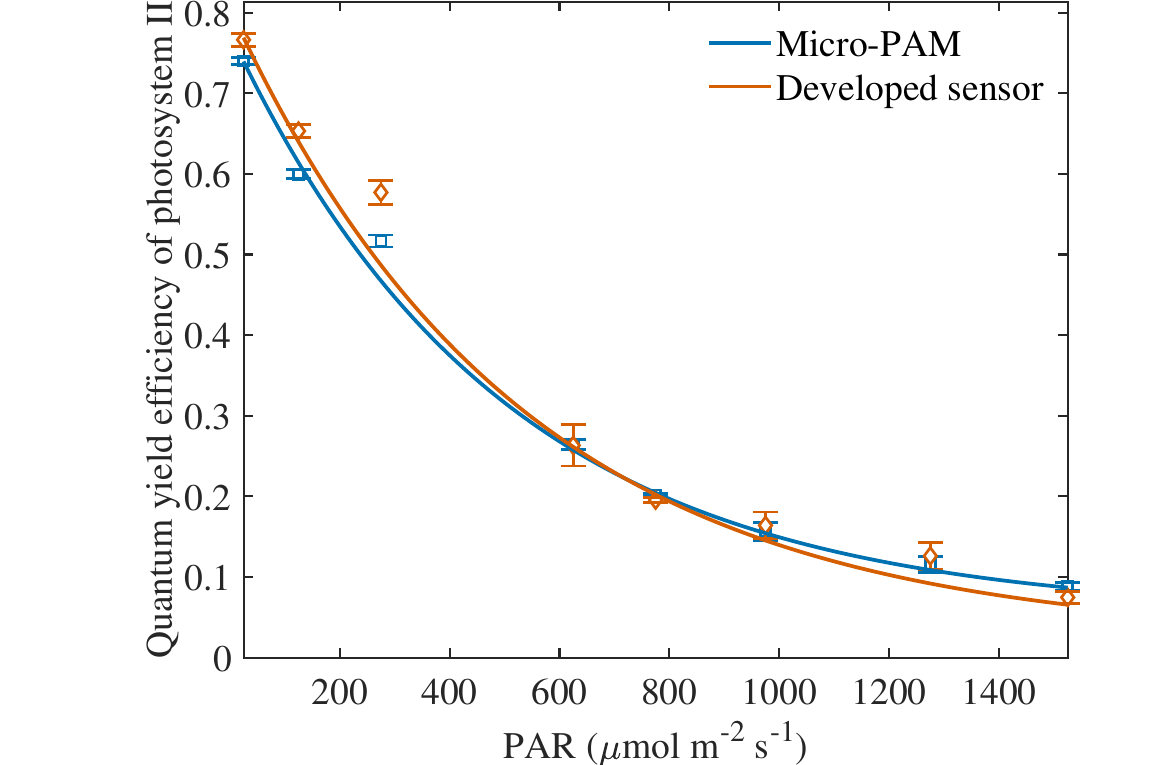}
    \caption{}
    \label{fig:Cherry_Light_Curve2}
  \end{subfigure}

  \vspace{3mm}

  \begin{subfigure}[b]{0.9\linewidth}
    \centering
    \includegraphics[width=\linewidth, trim=0mm 70mm 20mm 80mm, clip]{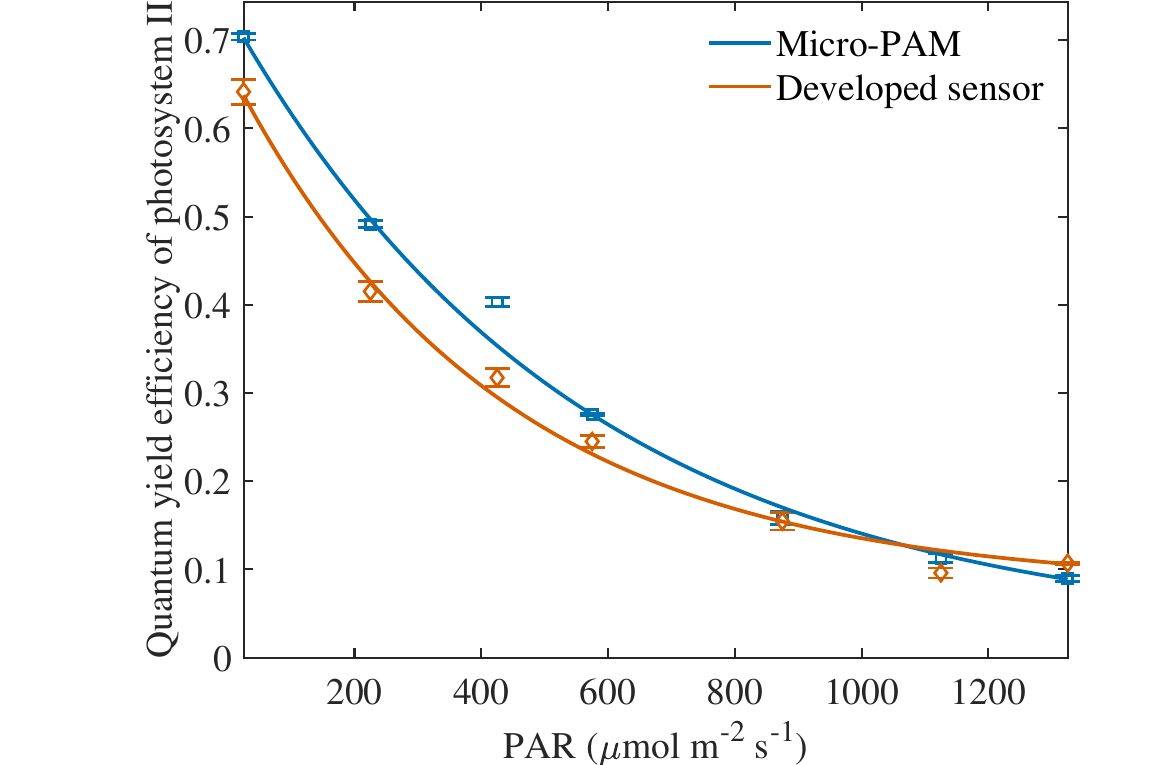}
    \caption{}
    \label{fig:Beech_Light_Curve2}

  \end{subfigure}

  \caption{Light-response curves from side-by-side measurements for three plant species: 
(a) sycamore maple, (b) cherry laurel, and (c) European beech. 
Each plot shows the quantum yield efficiency of photosystem II ($\Phi_\text{PSII}$) measured by the developed sensor and the commercial Micro-PAM under varying levels of ambient light. 
Data points represent the mean and standard deviation of at least three measurements. The fitted lines indicate the exponential decay trend for each device.
 }
  \label{fig:Light_Curve_All}
\end{figure}

Following the validation experiments described in the previous sections, the measurement results from the developed PAM fluorometer were compared with those obtained from the commercial sensor. 
Fig.~\ref{fig:Light_Curve_All} presents the quantum yield efficiency of PSII  measured by both instruments as a function of ambient photosynthetically active radiation (PAR) for the three tested species. 
Each data point represents the mean and standard deviation of at least three measurements at a constant PAR level. An exponential decay function was fitted to the data to describe the overall light response trend for both devices.

Across all species, both instruments showed the expected decrease in $\Phi_\text{PSII}$ with increasing PAR,  which reflects the photochemical quenching behavior. 
The developed sensor closely matched the Micro-PAM across the full range of light intensities and it reproduced both the shape and magnitude of the light response curves. 

For sycamore maple (Fig.~\ref{fig:Maple_Light_Curve2}), the developed and commercial sensors lead to nearly identical curves up to \SI{1000}{\micro\mole\per\square\meter\per\second}. 
Cherry laurel (Fig.~\ref{fig:Cherry_Light_Curve2}) showed strong correspondence across the entire range up to \SI{1500}{\micro\mole\per\square\meter\per\second}, while European beech (Fig.~\ref{fig:Beech_Light_Curve2}) also exhibited consistent tracking, with only slight deviations at intermediate intensities. 

Minor variations were observed at low light intensities, which may stem from the optical alignment and slightly different PAR values at the measurement points of the two sensors or from a small measurement delay between them. Since the sensors were positioned on the same leaf, it is possible that the burst saturation pulse from one sensor during measurement affected the other sensor results, as this deviation became more noticeable when the leaf surface area was smaller, such as in the European beech species. In addition, physiological differences or reflection effects at the leaf surface could have contributed to these deviations.

Overall, these results demonstrate that the developed sensor reliably reproduces light response curves across different plant species. 
The close agreement with the commercial Micro-PAM confirms that the new design achieves comparable measurement accuracy and is suitable for chlorophyll fluorescence assessments under natural light conditions.

\begin{figure}[!htbp]
  \centering
  \includegraphics[
    width=\linewidth,
    trim=0mm 50mm 0mm 60mm, clip
  ]{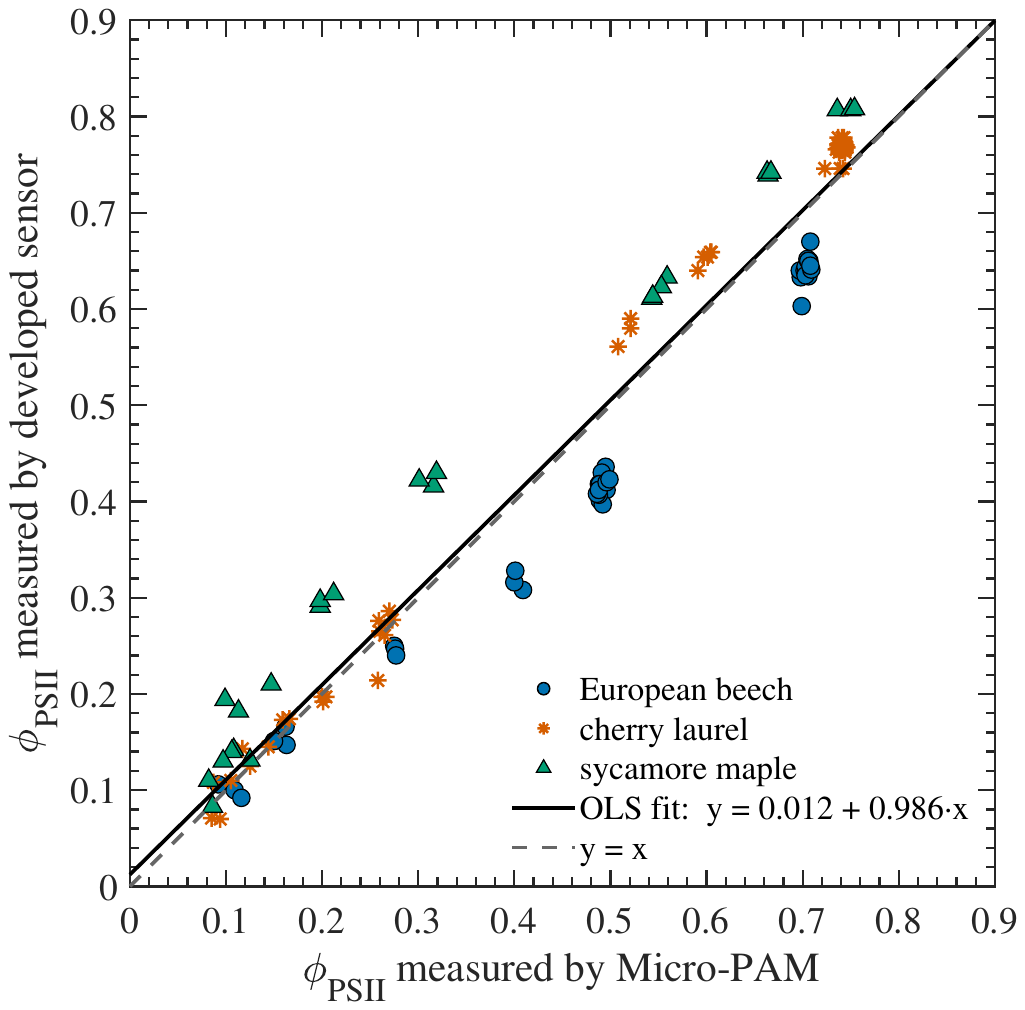} 
  \caption{Correlation between the quantum yield efficiency of photosystem II measured by the developed PAM fluorometer and the commercial Micro-PAM across all tested species and light intensities. 
The solid line represents the 1:1 relationship, and the dashed line indicates the ordinary least squares regression.
}
  \label{fig:Data_Correlation2}
\end{figure}

To further evaluate the correlation between the developed PAM fluorometer and the Micro-PAM, all quantum yield measurements obtained from the three species were combined in a single correlation analysis depicted in Fig.~\ref{fig:Data_Correlation2}. 
The data points from all species cluster closely around the 1:1 line, which indicates strong correlation between the two instruments. 
An ordinary least squares (OLS) regression resulted in a correlation coefficient of $R^2 = 0.95$. This confirms an excellent linear relationship across species and measurement conditions. 
The slope close to unity and negligible offset further verify that both instruments operate on a comparable scale and validates the measurement accuracy  of the newly developed sensor.

\subsection{Results of Sensor Reproducibility Tests}
\begin{figure}[!htbp]
  \centering
  \includegraphics[
    width=\linewidth,
    trim=0mm 50mm 0mm 60mm, clip
  ]{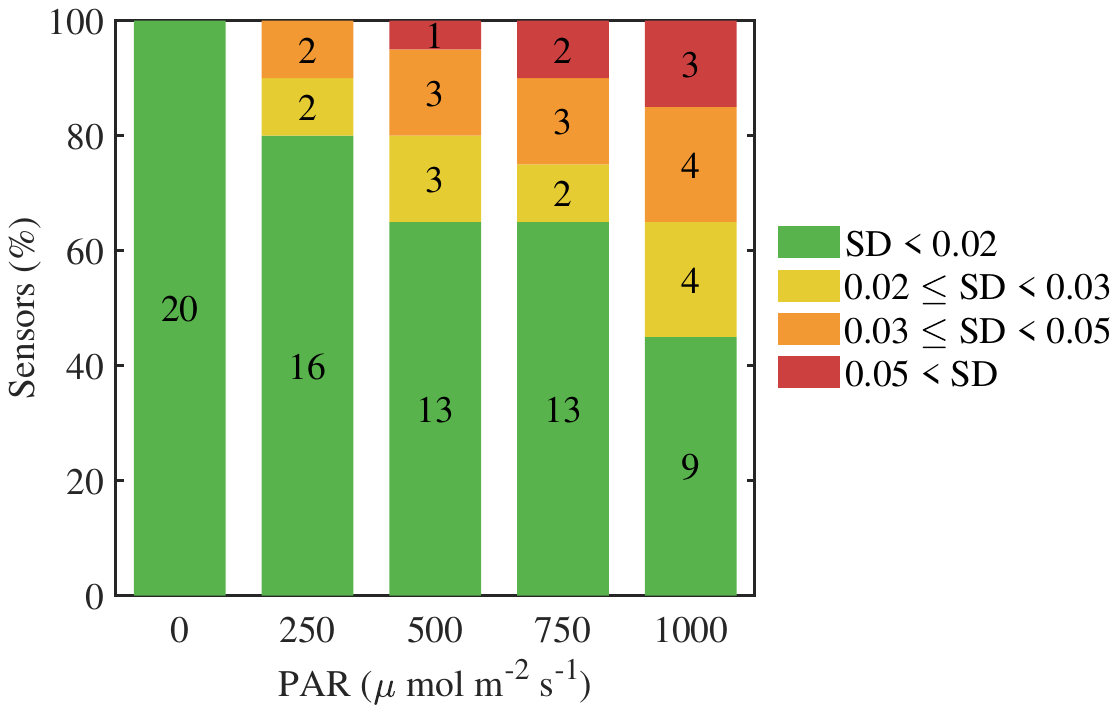} 
  \caption{Distribution of within-sensor standard deviation of $\Phi_\text{PSII}$ across PAR levels.
Each bar shows the percentage of sensors whose measurement variability falls within specific standard deviation intervals.
SD values represent the standard deviation (±) of repeated measurements for each sensor under identical light conditions.
}
  \label{fig:statistical results}
\end{figure}

We evaluated the measurement reproducibility of the developed PAM fluorometer under different light intensities using a batch of 20 sensors. 
For each sensor, measurements at every light intensity level were repeated at least five times to ensure sufficient data for calculating the mean and standard deviation (SD) of $\Phi_{\text{PSII}}$.
The SD at each light level was used as an indicator of measurement consistency. 
In this analysis, SD values below 0.02 were considered to represent good consistency, values between 0.02 and 0.03 moderate consistency, values between 0.03 and 0.05 reduced consistency, and values above 0.05 low measurement stability.
Fig.~\ref{fig:statistical results} shows how the proportion of sensors within these SD ranges changes with increasing PAR level.

Under dark conditions, all sensors featured good consistency in their measurements. As light intensity increased, some sensors showed reduced consistency and occasional fluctuations under strong illumination. 

Further investigation revealed that certain LEDs used as excitation sources delivered lower optical output at their fullwidth at half maximum (FWHM), and the saturation pulse intensity was not enough. Insufficient saturation pulse intensity under high light conditions led to incomplete PSII closure and, consequently, unstable or underestimated $F_m'$ values. This was also reported by Karageorgou et al.\ \cite{Karageorgou2007}.
 
Increasing the saturation pulse current effectively resolved this issue and improved measurement stability at high light levels. In the final tests, all 20 sensors achieved SD values below 0.03 across the full range of light intensities, which confirms stable and repeatable fluorescence measurements.

\section{Conclusion}

This contribution presents the development and validation of an autonomous low cost and compact PAM fluorometer designed for real-time, \textit{in situ} monitoring of photosynthetic efficiency in natural environments.

The prototype had cost of approximately 150~\euro{}, and was enclosed in a compact housing with geometric dimensions of \SI{3}{\centi\meter}~$\times$~\SI{6}{\centi\meter}~$\times$~\SI{2}{\centi\meter} and a weight of  \SI{50}{\gram}. With these specifications, it offers a scalable and lightweight solution for distributed deployment in forest ecosystems.

The developed sensor uses a single LED that alternates between measuring and saturation light modes. 
By modulating the saturation pulse with a higher frequency and duty cycle, measuring light pulses can be applied between the high intensity bursts and capture the maximum fluorescence before it drops until it reaches to saturation level. 
This configuration enables accurate monitoring of the fluorescence response during the saturation phase and avoids underestimation of $F_m'$ and $\Phi_{\text{PSII}}$, a limitation reported in previous sensor design by Haidekker et al.~\cite{haidekker2022lowcost}.

A key challenge reported in PAM fluorometer development lies in balancing sensitivity and dynamic range \cite{schreiber2004,haidekker2022lowcost}. 
The fluorescence signal induced by the weak measuring light (0.1~\si{\micro\mole\per\square\meter\per\second}) can be several orders of magnitude lower than the ambient light, which may reach up to 2000~\si{\micro\mole\per\square\meter\per\second} under full sunlight, or the saturation pulse, which can exceed 7,000~\si{\micro\mole\per\square\meter\per\second}. 
To address this, the analog front-end was designed for selective amplification and precise timing synchronization to isolate the fluorescence changes caused by the measuring pulse \cite{schreiber2004}. 

The prototype provides measurement results that were comparable to those obtained with a commercial PAM fluorometer at different levels of light intensity. Furthermore, reproducibility tests across a batch of sensors revealed that not all the sensors produce stable and consistent measurements under high light intensities. The source of these fluctuations were investigated and corrected. 
Moreover, further data processing and additional measurements are planned to determine whether individual sensors require calibration factors and how these should be applied.

Since the reproducibility tests were conducted over several months with the batch of sensors, it is currently unclear whether the observed measurement or light curve differences among sensors were caused by environmental factors and plant responses over time, or whether specific sensors require calibration. As there is no existing literature on this topic, no definitive conclusion can be drawn at this stage.

A batch of 20 sensors is currently being deployed at the ECOSENSE forest field site, where six Micro-PAM heads are also operating for long-term measurements.
This continuous monitoring of $\Phi_\text{PSII}$ throughout the vegetation season will provide valuable information on the durability, stability, and performance of the sensor under natural conditions. These PAM fluorometers are also equipped with PAR and temperature sensors developed within the ECOSENSE project. The miniaturized sensor head integrating these components has been developed by project partners and will be reported in a separate publication.

Several opportunities exist to further improve the developed PAM fluorometer. 
First, implementing adjustable intensity levels for both the measuring light and the saturation pulse would enable more flexible monitoring of $\Phi_\text{PSII}$ across different species and illumination conditions. 
 Second, integrating bidirectional LoRa communication will allow dynamic remote adjustment of excitation light intensities based on the collected data over time.

Using a plastic optical fiber to transfer the fluorescence was cost-effective. However, under harsh environmental conditions, such as wind and storms at the forest field site, it was prone to breakage. Therefore, we plan to slightly modify the design and remove the optical fiber. 

Future work will also focus on reducing power consumption and integrating energy harvesting solutions to eliminate the need for external power sources. 
Such improvements will minimize cabling, decrease overall sensor weight, and support a fully autonomous “deploy-and-forget” PAM fluorometer.

\section*{Acknowledgment}

The authors would like to thank Christoph Bohner for providing the sensor housings, and Eva Schottmüller for providing and taking care of the plants in the climate chamber. The authors also thank Dr. Simon Haberstroh for his help in analyzing the sensor data.

\bibliographystyle{ieeetr}
\bibliography{reference}

\end{document}